\documentstyle [12pt, axodraw] {article}

\catcode`@=12
\begin{document}
\thispagestyle{empty}
\begin{flushright} UCRHEP-T223\\MRI-PHY/P980444\\TIFR/TH/98-21\\June 1998\
\end{flushright}
\vspace{0.5in}
\begin{center}
{\Large	\bf Radiative Neutrino Mass Matrix for\\
Three Active plus One Sterile Species\\}
\vspace{.5in}
{\bf Naveen Gaur$^1$, Ambar Ghosal$^2$, Ernest Ma$^3$, and Probir Roy$^4$\\}
\vspace{0.3in}
{$^1$ \sl Department of Physics and Astrophysics,\\}
{\sl University of Delhi, Delhi 110 007, India \\} 
\vspace{0.1in}
{$^2$ \sl Mehta Research Institute, Chhatnag Road, Jhusi,\\}
{\sl Allahabad 211 019, India\\}
\vspace{0.1in}
{$^3$ \sl Department of Physics, University of California,\\}
{\sl Riverside, California 92521, USA\\}
\vspace{0.1in}
{$^4$ \sl Tata Institute of Fundamental Research,\\} 
{\sl Mumbai 400 005, India\\}
\vspace{.8in}
\end{center}
\begin{abstract}
A simple unifying mass matrix is presented for the three active and
one sterile neutrinos $\nu_e$, $\nu_\mu$, $\nu_\tau$, and $\nu_s$,
using an extension of the radiative mechanism proposed some time ago
by Zee.  The total neutrino-oscillation data are explained by the
scheme $\nu_e \leftrightarrow \nu_s$ (solar), $\nu_\mu \leftrightarrow
\nu_\tau$ (atmospheric) and $\nu_e \leftrightarrow \nu_\mu$ (LSND).
We obtain the interesting approximate relationship $(\Delta m^2)_{\rm
atm} \simeq 2 [(\Delta m^2)_{\rm solar} (\Delta m^2)_{\rm
LSND}]^{1/2}$ which is well satisfied by the data.
\end{abstract}

\newpage
\baselineskip 24pt

Three neutrinos, each associated with a charged lepton ($e$, $\mu$,
$\tau$), are now known.  The invisible width of the $Z$ boson, coming
from the decay $Z \to \nu \bar \nu$, is also consistent\cite{1} with
exactly three such neutrinos. This means that if there is a fourth
neutrino, either it has to be very heavy (with mass greater than
$M_Z/2$) or it does not couple to $Z$.  In particular, if it is light,
then it must not have any electroweak gauge interactions. Such an
object is often referred to as a ``sterile" neutrino.  The reason that
this may be a necessary part of our understanding of particle physics
is that there are at present three classes of neutrino
experiments\cite{2,3,4} which show evidence of neutrino oscillations
with three very different $\Delta m^2$'s, {\it i.e.} differences of
mass-squares.  If all three interpretations are correct, then we need
four light neutrinos.  (A possible but rather extreme three-neutrino
scenario\cite{5} is to have large anomalous $\nu_\tau$-quark
interactions.) It is thus of theoretical interest to find a natural
mechanism which explains the masses and mixings of these four
neutrinos in the present experimental context.

A specific model for a $4 \times 4$ neutrino mass matrix was
proposed\cite{6} already some time ago..  The form of this matrix
agrees with subsequent purely phenomenological analyses\cite{7,8} of
all neutrino-oscillation data.  Our present study concerns the
possibility that all neutrino masses are zero at tree level, but are
generated radiatively at one-loop to match the pattern in [6], using a
mechanism first proposed by Zee\cite{9}.  We extend previous
work\cite{10,11} on this topic to include a sterile neutrino\cite{12}
with the help of an extra U(1) gauge symmetry\cite{13}.  The resulting
mass eigenvalues lead to the approximate relationship
\begin{equation}
(\Delta m^2)_{\rm atm} \simeq 2 \sqrt {(\Delta m^2)_{\rm solar} 
(\Delta m^2)_{\rm LSND}}
\end{equation}
which is well satisfied by the data.

Our model extends the standard electroweak gauge model to
include three singlet fermion fields $\nu_{sL}$, $N_R$, and $S_R$, as well
as 3 singlet scalar fields $\chi_1^+$, $\chi_2^+$, and $\chi_2^0$.
There are also two scalar doublets $(\phi_1^+, \phi_1^0)$ and
$(\phi_2^+, \phi_2^0)$, where only one is needed in the minimal
standard model.  To obtain radiative masses for the three doublet
neutrinos, just $(\phi_{1,2}^+, \phi_{1,2}^0)$ and $\chi_1^+$ are
enough\cite{9,11}.  The more difficult task is to include the singlet
neutrino $\nu_{sL}$ into a $4 \times 4$ radiative mass matrix of the
same form.  A natural way that this may come about is to have an extra
gauge symmetry $U(1)'$ for the fields $\nu_{sL}$, $\chi_2^+$, and
$\chi_2^0$ which is broken at a higher $(\sim {\rm TeV})$ scale.  The
axial-vector anomaly, generated by $\nu_{sL}$, is 
cancelled by $N_R$ which transforms as $\nu_{sL}$ under $U(1)'$. We
also add $S_R$ which is trivial under $U(1)'$. A large mass for $N_R$
is then ensured through the Yukawa interaction $\bar S_RN_R{}^C
\chi^0_2$ since $\langle \chi^0_2 \rangle> \sim 1 \ {\rm TeV}$. The particle
content of the model is summarized in Table 1. 

\begin{table}
\begin{center}
\begin{tabular}{|c|c|c|c|}
\hline
fermions & L-parity & $SU(2)_L \times U(1)_Y$ & $U(1)'$ \\
\hline
&&&\\
$(\nu_i, l_i)_L$ & $-$ & $(2, -1/2)$ & 0 \\
$l_{iR}$ & $-$ & $(1,-1)$ & 0 \\
$\nu_{sL}$ & $-$ & (1,0) & 1 \\
$N_R$ & + & (1,0) & 1 \\
$S_R$ & + & (1,0) & 0 \\
&&&\\
\hline
scalars & L-parity & $SU(2)_L \times U(1)_Y$ & $U(1)'$ \\
\hline
&&&\\
$(\phi_{1,2}^+,\phi_{1,2}^0)$ & + & (2,1/2) & 0 \\
$\chi_1^+$ & + & (1,1) & 0 \\
$\chi_2^+$ & + & (1,1) & 1 \\
$\chi_2^0$ & + & (1,0) & 1 \\
&&&\\
\hline
\end{tabular}
\caption{List of fermion and scalar fields in our model.}
\end{center}
\end{table}

We have an unbroken discrete $Z_2$ symmetry, namely L-parity, to
distinguish between two classes of fermions.  The leptons now have odd
L-parity, replacing the usual additive lepton number.  This allows the
four neutrinos to acquire Majorana masses.  However, tree-level
neutrino masses are forbidden by the assumed particle content of our
model, even after the spontaneous breaking of the gauge symmetry.
Note that $\nu_s$ does not get a Majorana mass because of $U(1)'$; it
also does not get a Dirac mass by pairing up with $N_R$ or $S_R$
because of L-parity.  More specifically, consider the following
interaction Lagrangian density of the fields shown in Table 1.
\begin{eqnarray}
{\cal L}_{int} &=& \sum_{i,j} f_{ij} (\nu_{iL} l_{jL} - l_{iL} \nu_{jL}) 
\chi_1^+   + \sum_i f'_i \bar\nu_{sL} l_{iR} \chi_2^+   + 
\sum_i h_i (\bar\nu_{iL} \phi_1^+ + \bar l_{iL} \phi_1^0) 
l_{iR} \nonumber \\
&& + \mu (\phi_1^+ \phi_2^0 - \phi_1^0 \phi_2^+) \chi_1^-   + 
\mu' \chi_1^+ \chi_2^- \chi_2^0 + h' N_R S_R {\chi_2^0}^\ast + h.c.,
\end{eqnarray}
where we have used the notation $\psi_i \zeta_j = \overline{\psi_i{}^C}
\zeta_j$ for two fermion fields $\psi$ and $\zeta$. Evidently, $f_{ij}$
is antisymmetric in its generation indices. We have assumed in (2)
that $(\phi^+_2, \phi^0_2)$ do not couple to leptons.  This is easily
achieved by a separate discrete $Z_2$ symmetry which is explicitly
broken, but only by soft terms such as $\phi^-_1\phi^+_2 + \phi^{0
\ast}_1 \phi^0_2 + h.c.$ in the Higgs potential, as in the minimal
supersymmetric standard model, for example.  As shown below, the above
interactions induce a radiative neutrino mass matrix for $\nu_e$,
$\nu_\mu$, $\nu_\tau$, and $\nu_s$ of the form
\begin{equation}
{\cal M}_\nu = \left[ \begin{array} {c@{\quad}c@{\quad}c@{\quad}c} 0 & a & b & 
d \\ a & 0 & c & e \\ b & c & 0 & f \\ d & e & f & 0 \end{array} \right],
\end{equation}
which generalizes the $3 \times 3$ matrix of the Zee model [9] by
including a fourth row and column.

In Fig.~1 we show the one-loop diagram linking $\nu_i$ and $\nu_j$
which contributes to the corresponding entry in ${\cal M}_\nu$.  This
is of course identical to that of Ref.~[9] and [11].  Note that $i
\neq j$ necessarily, hence only off-diagonal entries can be nonzero.
Since $h_i = m_{l_i} / \langle \phi_1^0 \rangle$, we obtain
\begin{eqnarray}
a &=& f_{e \mu} (m_\mu^2 - m_e^2) \left( {\mu v_2 \over v_1} \right) 
F(m_{\chi_1}^2, m_{\phi_1}^2), \\ 
b &=& f_{e \tau} (m_\tau^2 - m_e^2) \left( {\mu v_2 \over v_1} \right) 
F(m_{\chi_1}^2, m_{\phi_1}^2), \\
c &=& f_{\mu \tau} (m_\tau^2 - m_\mu^2) \left( {\mu v_2 \over v_1}
\right)  F(m_{\chi_1}^2, m_{\phi_1}^2),
\end{eqnarray} 
where $v_{1,2} \equiv \langle \phi^0_{1,2} \rangle$, and the function $F$ 
is given by
\begin{equation}
F(m_1^2, m_2^2) = {1 \over 16 \pi^2} {1 \over m_1^2 - m_2^2} \ln {m_1^2 
\over m_2^2}.
\end{equation}

In Fig.~2 we show the analogous one-loop diagram linking $\nu_i$ to
$\nu_s$.  We find
\begin{eqnarray}
d &=& (f_{e \tau} f'_\tau m_\tau + f_{e \mu} f'_\mu m_\mu) 
\left( {\mu' u \over v_1} \right) F(m_{\chi_1}^2, m_{\chi_2}^2), \\
e &=& (f_{\mu \tau} f'_\tau m_\tau + f_{\mu e} f'_e m_e) 
\left( {\mu' u \over v_1} \right) F(m_{\chi_1}^2, m_{\chi_2}^2), \\
f &=& (f_{\tau \mu} f'_\mu m_\mu + f_{\tau e} f'_e m_e) 
\left( {\mu' u \over v_1} \right) F(m_{\chi_1}^2, m_{\chi_2}^2),
\end{eqnarray}
where $u \equiv \langle \chi_2^0 \rangle$.  In the following, we will
assume that $f'_e m_e$ is negligible.  Moreover, while $u$ is expected
to be large compared to $v_{1,2}$, that can be compensated by $m_{\chi_2}$
being larger than $m_{\chi_1}$ or $m_{\phi_1}$.  Thus 
$d,e,f$ need not be larger in magnitude than $a,b,c$.  In any case, we have 
the important relationship
\begin{equation}
d  = {b e\over c} \left( 1 - {m_\mu^2 \over m_\tau^2} \right)  +
{ff_{e \mu} \over f_{\tau \mu}},
\end{equation}
where $m_e^2$ in Eq.~(5) has been neglected.

We make the same observation as in Refs.~[9] and [11] that $b$ and $c$
are likely to be the dominant entries of ${\cal M}_\nu$ because they
are proportional to $m_\tau^2$.  This means that $\nu_\tau$ combines
with a linear combination of $\nu_e$ and $\nu_\mu$ to form a
pseudo-Dirac particle.  Let us also assume that $|f_{e \tau}| << |f_{\mu
\tau}|$, so that $|b| << |c|$. Then the $2 \times 2$ submatrix spanning
$\nu_e$ and $\nu_s$ is given by
\begin{eqnarray}
{\cal M}_{\nu_e \nu_s} &=& \left[ \begin{array} {c@{\quad}c} -2ab/c & 
d - be/c - af/c \\ d - be/c - af/c & -2ef/c \end{array} \right]
\nonumber \\  
&=& \left[ \begin{array} {c@{\quad}c} -2ab/c & -ff_{e \mu}/f_{\mu \tau} - 
(be/c)(m_\mu^2/m_\tau^2) \\ -ff_{e \mu}/f_{\mu \tau} - (be/c)(m_\mu^2/
m_\tau^2) & -2ef/c \end{array} \right],
\end{eqnarray}
where we have used Eq.~(11) and the fact that $|a/c| << |f_{e \mu}/f_{\mu
\tau}|$.  Hence
\begin{equation}
m_{\nu_e} \simeq -2 {ab \over c}, ~~~ m_{\nu_s} \simeq -2 {ef \over c},
\end{equation}
and for $m_{\nu_e} << m_{\nu_s}$, the $\nu_e - \nu_s$ mixing is
$(cf_{e \mu}/ef_{\mu \tau} + b m_\mu^2/f m_\tau^2)/2$.  This is
assumed to be small, so as to satisfy the solar neutrino data.  We now
have
\begin{equation}
(\Delta m^2)_{\rm solar} \simeq 4 {e^2 f^2 \over c^2}.
\end{equation}

Since ${\cal M}_\nu$ has zero trace, it can easily be shown that the
leading expressions for its eigenvalues are given by
\begin{equation}
-2 {ab \over c}, ~~~ c + {ab \over c} + {ef \over c}, ~~~ -c + {ab
\over c} + {ef \over c}, ~~~ -2 {ef \over c}.
\end{equation}
Hence the mass-squared difference between the two Majorana components 
of the pseudo-Dirac neutrino with mass $c$ is
\begin{equation}
\Delta m^2 = 4 (ab + ef) \simeq 4 ef \simeq (\Delta m^2)_{\rm atm}.
\end{equation}
Since this is for a $\nu_\mu - \nu_\tau$ mixing of 45$^\circ$, we have 
taken it to explain the atmospheric neutrino data.  Finally, the LSND 
data involve the mixing of $\nu_e$ and $\nu_\mu$, hence
\begin{equation}
(\Delta m^2)_{\rm LSND} = c^2,
\end{equation}
with mixing given by $b/c$.  Combining Eqs.~(14), (16), and (17), we 
obtain Eq.~(1), as claimed.

Current neutrino-oscillation data are consistent with $(\Delta
m^2)_{\rm LSND} \sim 1~{\rm eV}^2$ and $(\Delta m^2)_{\rm solar}$
$\sim 6 \times 10^{-6}~ {\rm eV}^2$.  In that case, $(\Delta m^2)_{\rm
atm}$ is predicted by Eq.~(1) to be about $5 \times 10^{-3}~{\rm
eV}^2$, which is supported by the most recent data from
Super-Kamiokande.  In our model, $\nu_\mu$ and $\nu_\tau$ have the
same mass $c \simeq 1$ eV and they mix maximally.  Let $b \simeq 0.04$
eV, then the $\nu_\mu - \nu_e$ mixing parameter $(\sin^2 2
\theta)_{\rm LSND}$ is $4 b^2/c^2 \sim 6 \times 10^{-3}$, in good
agreement with data.  For $\nu_e - \nu_s$ oscillations, we let
\begin{equation}
{f_{e \mu}c \over 2f_{\mu \tau}e} + {b m_\mu^2 \over 2 f m_\tau^2}
\simeq 0.04,  
\end{equation}
so that $(\sin^2 2 \theta)_{\rm solar}$ is also about $6 \times
10^{-3}$, again in good agreement with data.  More specifically, we
can let $e \simeq 0.12$ eV and $f \simeq 0.01$ eV, then $m_{\nu_s}
\sim 2ef/c \simeq 2.4 \times 10^{-3}$ eV.  Furthermore from Eq.~(18),
$f_{e \mu}/f_{\mu \tau}$ is now about 0.008 and from Eqs.~(4) and (6),
$a \sim 3 \times 10^{-5}$ eV, hence $m_{\nu_e} \sim 2ab/c \sim 2
\times 10^{-6}$ eV, justifying our assumption that $m_{\nu_e} <<
m_{\nu_s}$.  We have thus a completely successful phenomenological
picture of neutrino oscillations.

The model of Ref.[11] differs from ours in that $\nu_s$ is assumed there 
to acquire a tree-level mass which is just slightly bigger than the 
radiative mass of $\nu_e$.  [This is of course rather {\it ad hoc}, but it 
is necessary to satisfy solar data.]  Let us compare its consequences with 
those of ours. In the former, the parameter $a$ is forced to be large in
magnitude because $4ab$ is identified there with $(\Delta m^2)_{\rm
atm}$, resulting in $|f_{e \mu}| \sim |f_{\mu \tau}|$.  This condition
is subject to severe phenomenological constraints because $f_{e \mu}$
contributes to $\mu$ decay. In fact, in that scenario, $f^2_{e\mu}
\sim f^2_{\mu\tau} < 7 \times 10^{-4} G_F$. $(\cos^2\phi M^{-2}_1 +
\sin^2\phi M^{-2}_2)^{-1}$ where $M_{1,2}$ are the physical charged
Higgs masses and $\phi$ is their mixing angle.  In our model, because
of Eq.~(16), $a$ can be and is very small, hence $|f_{e \mu}| <<
|f_{\mu \tau}|$, so that our $|f_{\mu\tau}|$ is not constrained to be
small.

We note also that the form of Eq.~(3) for the neutrino mass matrix
with $c$ as the dominant entry is not sufficient by itself to have the
correct $\nu_e - \nu_s$ submatrix needed to explain the solar data.
Without Eq.~(11), which is an \underline {automatic} consequence of our
model, that submatrix would have dominant off-diagonal terms, {\it
i.e.}
\begin{equation}
{\cal M}_{\nu_e \nu_s} \sim \left( \begin{array} {c@{\quad}c} 0 & d \\
d & 0  
\end{array} \right),
\end{equation}
which would make $\nu_e$ and $\nu_s$ pseudo-Dirac partners with the
requisite mixing of 45$^\circ$ in conflict with solar neutrino data.

A third point concerns the fermion singlets $N_R$ and $S_R$.  They
have even L-parity, which is unbroken in our model, hence they do not
mix into the lepton sector.  Both of them are massive, because the
terms $S_R S_R$ and $N_R S_R {\chi_2^0}^\ast + h.c.$ are allowed in
the Lagrangian density. The scale of $U(1)'$-breaking, i.e. $\langle
\chi^0_2\rangle$ can be taken beyond 1 TeV, thereby pushing up these
masses. It is to be noted that the off-diagonal terms in the $Z -
Z^\prime$ mass matrix are prohibited due to the absence of
appropriate Higgs fields in the present model.  We also assume that
the kinetic mixing between the $U(1)_Y$ and $U(1)'$ gauge bosons is
negligible.  Hence our $Z'$ couples at the tree level only to
$\nu_{sL}$, $N_R$, $\chi^+_2$ and $\chi^0_2$.  Thus present
experimental bounds [14] on a possible $Z'$ with standard-model-like
couplings do not apply.  However, because $\nu_s$ mixes with $\nu_e$
radiatively, $Z'$ develops a small coupling to $\nu_e$.  To avoid any
possible conflict with nucleosynthesis or current electroweak
phenomenology, we assume $M_Z \sim 1$ TeV or greater, which is of
course natural since we already assumed $\langle \chi^0_2 \rangle \sim
1$ TeV or greater.

The charged scalar $\chi_1^+$ contributes to the standard-model
effective charged-current interaction due to the presence of the
$f_{ij}(\nu_{iL} l_{jL} - l_{iL} \nu_{jL})\chi_1^+$ term in the
Lagrangian density. The corresponding effects on processes, such as
electron-neutrino scattering, are experimentally severely constrained.
They give rise to the constraint $f_{e\mu}^2/M^2 < 0.036 G_F$
[11], where $M$ is the mass of the charged scalar mediating the
process.  Since we have $|f_{e\mu}| << |f_{\mu\tau}|$, this is no
problem for us. The proposed hierarchical relation
$|f_{e\mu}|<<|f_{\mu\tau}|$ is also consistent with the constraint
from the branching ratio of the decay 
$\tau^- \rightarrow\mu^-\bar \nu_\mu\nu_\tau$
being $(17.35 \pm 0.10)\%$
\cite{14} since the latter only requires $f^2_{\mu\tau}/M^2 < 0.13 \
G_F$.  

In summary, we have demonstrated that the present results of solar,
atmospheric as well as LSND experiments can be explained with three
electroweak-active neutrinos and a sterile one with a minimal 
extension of the standard  $SU(2)_L\times U(1)_Y$ electroweak gauge 
model. The extra $U(1)^\prime$ gauge and $Z_2$ discrete symmetries are
needed to avoid tree-level Majorana or Dirac mass terms. All neutrino
masses are radiatively generated in one loop by an extension of the
Zee model. Our proposal results in an interesting 
relationship $(\Delta m^2)_{\rm{atm}}\simeq 2{[(\Delta m^2)_{\rm{solar}} 
(\Delta m^2)_{\rm{LSND}}]}^{1/2}$ which is well satisfied by the
present experimental data and will be critically tested with more
accurate data forthcoming in the near future. 
\vspace{0.3in}
\begin{center} {ACKNOWLEDGEMENT}
\end{center}

The authors acknowledge the hospitality and 
stimulating environment of WHEPP 5 (Fifth Workshop on High Energy
Physics Phenomenology), held at IUCAA, Pune, India under the auspices
of the S.N. Bose National Centre for Basic Sciences and the Tata
Institute of Fundamental Research, where this
study was initiated.  The work of E.M. was supported in part by the
U.~S.~Department of Energy under Grant No.~DE-FG03-94ER40837.

\newpage

\bibliographystyle{unsrt}

\begin{thebibliography}{99}
\bibitem{12} In Ref.[11], the sterile neutrino has a tree-level mass,
while all our neutrino masses are of radiative origin.
\end{thebibliography}

\newpage
%
\vskip 1in
\begin{center}
\begin{picture}(200,100)(-0,-0)
\ArrowLine(0,0)(60,0)
       \Text(30,-8)[c]{$\nu_{iL}$}   
\DashArrowArc(100,0)(40,0 ,90)6
       \Text(59,28)[c]{$\chi_1^+$}  
\ArrowLine(100,0)(60,0)
       \Text(80,-8)[c]{$l_{jL}$}  
\DashLine(100,0)(100,-40)6
       \Text(100,-47)[c]{$<\phi_1^0>$}
\ArrowLine(140,0)(100,0)
       \Text(120,-8)[c]{$l_{jR}$}
\ArrowLine(200,0)(140,0)
       \Text(180,-8)[c]{$\nu_{jL}$}
\DashArrowArc(100,0)(40,90 ,180)6
       \Text(142,28)[c]{$\phi_1^+$}  
\DashLine(100,40)(100,80)6
       \Text(100,87)[c]{$<\phi_2^0>$}
\end{picture}
\end{center}
\vskip 1in
\begin{center}
{\bf{FIG. 1. }}One loop radiative $\nu_i - \nu_j$ ($i,j$ = $e,\mu,\tau$) 
mass due to charged Higgs exchange.
\end{center}
\vskip 1in
\begin{center}
\begin{picture}(200,100)(-0,-0)
\ArrowLine(0,0)(60,0)
       \Text(30,-8)[c]{$\nu_{iL}$}   
\DashArrowArc(100,0)(40,0 ,90)6
       \Text(59,28)[c]{$\chi_1^+$}  
\ArrowLine(100,0)(60,0)
       \Text(80,-8)[c]{$l_{jL}$}  
\DashLine(100,0)(100,-40)6
       \Text(100,-47)[c]{$<\phi_1^0>$}
\ArrowLine(140,0)(100,0)
       \Text(120,-8)[c]{$l_{jR}$}
\ArrowLine(200,0)(140,0)
       \Text(180,-8)[c]{$\nu_{sL}$}
\DashArrowArc(100,0)(40,90 ,180)6
       \Text(142,28)[c]{$\chi_2^+$}  
\DashLine(100,40)(100,80)6
       \Text(100,87)[c]{$<\chi_2^0>$}
\end{picture}
\end{center}
\vskip 1in
\begin{center}
{\bf{FIG. 2.}} One loop radiative $\nu_i- \nu_s$ ($i = e, \mu, \tau$)
mass due to charged Higgs exchange.
\end{center}
\end{document}